# How to Introduce Time Operator

Zhi-Yong Wang*, Cai-Dong Xiong

*School of Physical Electronics, University of Electronic Science and Technology of China, Chengdu 610054*

**Abstract**

Time operator can be introduced by three different approaches: by pertaining it to dynamical variables; by quantizing the classical expression of time; taken as the restriction of energy shift generator to the Hilbert space of a physical system.

*PACS*: 03.65.Ca; 03.65.Ta; 03.65.Xp

*Keywords*: time operator; self-adjointness; energy shift; time-energy uncertainty

## 1. Introduction

Traditionally, time enters quantum mechanics as a parameter rather than a dynamical operator. As a consequence, the investigations on tunneling time, arrival time and traversal time, etc., still remain controversial today [1-19]. On the one hand, one imposes self-adjointness as a requirement for any observable; on the other hand, according to Pauli's argument [20-23], there is no self-adjoint time operator canonically conjugating to a Hamiltonian if the Hamiltonian spectrum is bounded from below. A way out of this dilemma set by Pauli's objection is based on the use of positive operator valued measures (POVMs) [19, 22-26]: quantum observables are generally positive operator valued measures, e.g., quantum observables are extended to maximally symmetric but not necessarily self-adjoint operators, in such a way one preserves the requirement that time operator be conjugate to the Hamiltonian but abandons the self-adjointness of time operator.

In this paper, general time operators are constructed by three different approaches. In the following, the natural units of measurement ($\hbar = c = 1$) is applied.

---

* *E-mail address*:　zywang@uestc.edu.cn

## 2. Mandelstam–Tamm version of non-self-adjoint time operator

The time-energy uncertainty relation has been a controversial issue and has many versions. However, the Mandelstam–Tamm version [27] of the time-energy uncertainty is the most widely accepted nowadays, where the time deviation $\Delta T$ is given by a characteristic time associated with some dynamical variables. Similarly, we will show that time operator can be introduced by pertaining it to dynamical variables. Let $\hat{F}$ be a non-stationary observable (but does not depend explicitly on time $t$: $\partial \hat{F}/\partial t = 0$), in Heisenberg picture it satisfies the Heisenberg equation of motion for $\hat{F}$

$$\mathrm{i}\,\mathrm{d}\hat{F}/\mathrm{d}t = [\hat{F}, \hat{H}]. \tag{1}$$

Assume that whenever $\mathrm{d}\hat{F}/\mathrm{d}t$ is nonzero such that it has the inverse

$$(\mathrm{d}\hat{F}/\mathrm{d}t)^{-1} = \mathrm{i}([\hat{F}, \hat{H}])^{-1}. \tag{2}$$

If $[(\mathrm{d}\hat{F}/\mathrm{d}t)^{-1}, \hat{H}] = 0$, i.e., $[([\hat{F}, \hat{H}])^{-1}, \hat{H}] = 0$, one can introduce a time operator $\hat{T}_1$ as follows:

$$\begin{aligned} \hat{T}_1 &= -[\hat{F}(\mathrm{d}\hat{F}/\mathrm{d}t)^{-1} + (\mathrm{d}\hat{F}/\mathrm{d}t)^{-1}\hat{F}]/2 \\ &= -\mathrm{i}[\hat{F}([\hat{F}, \hat{H}])^{-1} + ([\hat{F}, \hat{H}])^{-1}\hat{F}]/2 \end{aligned}. \tag{3}$$

Using Eqs. (1)-(3) and $[([\hat{F}, \hat{H}])^{-1}, \hat{H}] = 0$, one has

$$[\hat{H}, \hat{T}_1] = \mathrm{i}, \tag{4}$$

i.e., $\hat{T}_1$ and $\hat{H}$ form a canonically conjugate pair. Owing to the fact that the Hamiltonian spectrum is bounded from below, the time operator $\hat{T}_1$ satisfying Eq. (4) is not self-adjoint. However, according to the formalism of POVMs [21-26], it can represent an observable.

As an example, assume that in (1+1) dimensional space-time ($t$, $x$), a freely moving particle (with mass $m$) has position $x$ and momentum $\hat{p} = -\mathrm{i}\,\partial/\partial x$, and its the Hamiltonian is denoted as $\hat{H} = \hat{p}^2/2m$. Let $\hat{F} = x$, applying Eq. (3) and $[x, \hat{p}] = \mathrm{i}$, one has

$$\hat{T}_1 = -m(\hat{p}^{-1}x + x\hat{p}^{-1})/2, \tag{5}$$

or in the momentum representation,

$$\hat{T}_1 = -\frac{\mathrm{i}m}{2}(\frac{1}{p}\frac{\partial}{\partial p} + \frac{\partial}{\partial p}\frac{1}{p}), \tag{6}$$

it is the free-motion time-of-arrival operator that has been studied in many literatures (see for example, Ref.[11, 19, 21-26]). For the time being, the time operator is defined on a dense domain of the Hilbert space of square integrable functions on the half real-line [11, 19, 21-26], denoted as $\mathcal{D}(\hat{T}) = \mathcal{H}_0 = L^2(\mathcal{R}^+, \mathrm{d}E)$, where $E$ corresponds to the eigenvalue of $\hat{H} = \hat{p}^2/2m$ and $\mathcal{R}^+ = \{E \mid 0 \leq E < +\infty\}$.

## 3. Mandelstam–Tamm version of self-adjoint time operator

In addition to the non-self-adjoint time operator $\hat{T}_1$ introduced by pertaining it to dynamical variables, based on the Mandelstam–Tamm version of the time-energy uncertainty, one can also introduce a self-adjoint time operator (say, $\hat{T}_2$). For this purpose, let $\hat{A}$ be a non-stationary observable (but does not depend explicitly on time $t$: $\partial \hat{A}/\partial t = 0$), in Heisenberg picture it satisfies the Heisenberg equation of motion for $\hat{A}$

$$\mathrm{i}\,\mathrm{d}\hat{A}/\mathrm{d}t = [\hat{A}, \hat{H}]. \tag{7}$$

Assume that whenever $\mathrm{d}\hat{A}/\mathrm{d}t$ is nonzero such that it has the inverse $(\mathrm{d}\hat{A}/\mathrm{d}t)^{-1} = \mathrm{i}([\hat{A}, \hat{H}])^{-1}$. Let $<\hat{A}>$ denote the quantum mechanical average of $\hat{A}$ (and so on), and then the variance of the $\hat{A}$ distribution is

$$\varDelta A = \sqrt{<(\hat{A} - <\hat{A}>)^2>}, \tag{8}$$

and so on. The Mandelstam–Tamm version [27] of the time-energy uncertainty is

$$\varDelta T_2 \varDelta H \geq 1/2, \tag{9}$$

where

$$\Delta T_2 = \Delta A / |< d\hat{A}/dt >|. \tag{10}$$

On the other hand, similar to Eq. (8), that is, according to the definition of quantum mechanical variance, one has

$$\Delta T_2 = \sqrt{< (\hat{T}_2 - < \hat{T}_2 >)^2 >}. \tag{11}$$

Comparing Eqs. (11) with (10), a time operator $\hat{T}_2$ is introduced

$$\hat{T}_2 = -\hat{A}/< d\hat{A}/dt >. \tag{12}$$

In fact, using Eqs. (8) and (11)-(12), one can easily obtain Eq. (10). In contrast with $\hat{T}_2$, the time operator $\hat{T}_1$ generally does not satisfy Eq. (10).

Because $\hat{A}$ and $\hat{H}$ are self-adjoint operators, and $< d\hat{A}/dt > = -i < [\hat{A}, \hat{H}] >$ is a real c-number, the time operator $\hat{T}_2$ is a self-adjoint one. At the price of this, one can show that the canonical commutation relation between $\hat{T}_2$ and $\hat{H}$ holds only in the sense of quantum-mechanical average

$$<[\hat{H}, \hat{T}_2]> = i, \tag{13}$$

which is similar to the Gupa-Bleuler formalism of Lorentz gauge quantization of electromagnetic field (see for example, Ref. [28]), where assume that the expectation value of $\partial_\mu A^\mu$, rather than $\partial_\mu A^\mu$ itself, vanishes: $<\partial_\mu A^\mu> = 0$ ($A^\mu$ ($\mu = 0,1,2,3$) represents the electromagnetic 4-potential).

As an example, let us consider the standard example of a particle in a constant field [22, 29], e.g., in the case of a freely falling particle in a homogeneous force filed. The Hamiltonian is $\hat{H} = \hat{p}^2/2m + mg\hat{q}$, where $\hat{q}$ denotes the position operator satisfying $\partial \hat{q}/\partial t = 0$ and $[\hat{q}, \hat{p}] = i$, one has

$$< \mathrm{d}\hat{p}/\mathrm{d}t > = -\mathrm{i} < [\hat{p}, \hat{H}] > = -\mathrm{i} < -\mathrm{i}mg > = -mg. \tag{14}$$

Now choose $\hat{A} = \hat{p}$ as the dynamical variable, via Eq. (12) the time operator $\hat{T}_2$ is

$$\hat{T}_2 = -\hat{A}/<\mathrm{d}\hat{A}/\mathrm{d}t> = \hat{p}/mg, \tag{15}$$

which can be ascribed to the time of arrival to zero momentum rather than to a specified position. In the present case, it is easy to show that $<[\hat{H},\hat{T}_2]> = [\hat{H},\hat{T}_2] = \mathrm{i}$, i.e., the commutation relation between $\hat{T}_2$ and $\hat{H}$ also holds in the usual sense. Therefore, $\hat{T}_2$ will do as a self-adjoint operator canonically conjugate to the Hamiltonian $\hat{H}$, which does not contradict Pauli's theorem because the present Hamiltonian is unbounded.

## 4. Time operator derived from quantizing the classical expression of time

A natural way of introducing time operator is based on the usual quantization procedure. As an example, let us consider the arrival time of free particles in (1+1) dimensional space-time. The classical expression for the arrival time at the origin $x_0=0$ of a freely moving particle having position $x$ and uniform velocity $v$, is $t = -x/v$. In a nonrelativistic theory, the particle of rest mass $m$ has momentum $p = mv \neq 0$ such that $t = -mx/p$. The transition from the classical expression to a quantum-mechanical description requires us to replace all dynamical variables with the corresponding linear operators, and symmetrize the classical expression. In this manner, one can obtain the nonrelativistic free-motion time-of-arrival operator ($\hat{p} = -\mathrm{i}\partial/\partial x$)

$$\hat{T}_{\mathrm{non}} = \hat{T}_{\mathrm{non}}(x, \hat{p}) = -m(\hat{p}^{-1}x + x\hat{p}^{-1})/2. \tag{16}$$

Using the Hamiltonian $\hat{H} = \hat{p}^2/2m$ and $[x, \hat{p}] = \mathrm{i}$, one can easy show that $[\hat{T}_{\mathrm{non}}, \hat{H}] = -\mathrm{i}$. The nonrelativistic time-of-arrival operator given by Eq. (16) has been studied thoroughly in previous literatures [11, 19, 21, 22].

In a similar manner, one can introduce a relativistic free-motion time-of-arrival operator.

In the relativistic case, in the natural units of measurement ($\hbar = c = 1$), the classical expression for the arrival time at the origin $x_0$=0 of a freely moving particle having position $x$ and uniform velocity $v$, is $t$=-$x$/$v$=-$x(E/p)$, where the minus in $t$=-$x(E/p)$ describes the past, $E$ is the relativistic energy of the particle satisfying $E^2 = p^2 + m^2$. Replacing all dynamical variables with the corresponding linear operators, and symmetrize the classical expression $t = -x/v = -Ex/p$, one can obtain the relativistic time operator as follows:

$$\hat{T}(x,\hat{p}) = -(1/6)[\hat{H}(\hat{p}^{-1}x + x\hat{p}^{-1}) + (\hat{p}^{-1}x + x\hat{p}^{-1})\hat{H} + \hat{p}^{-1}\hat{H}x + x\hat{H}\hat{p}^{-1}]. \quad (17)$$

In the momentum representation, the Hamiltonian is denoted as $H$, Eq. (17) becomes

$$\hat{T}(\hat{x},p) = -\frac{i}{6}[H(\frac{1}{p}\frac{\partial}{\partial p} + \frac{\partial}{\partial p}\frac{1}{p}) + (\frac{1}{p}\frac{\partial}{\partial p} + \frac{\partial}{\partial p}\frac{1}{p})H + \frac{1}{p}H\frac{\partial}{\partial p} + \frac{\partial}{\partial p}H\frac{1}{p}]. \quad (18)$$

For our purpose, let us assume that whenever $p \neq 0$, i.e., $E^2 \neq m^2$.

Firstly, as a simple example, let us consider the free spin-0 particles. For the moment, for simplicity let $H = E$, using $\partial/\partial p = p\partial/E\partial E$ and rewriting $\hat{T}(\hat{x},p)$ as $\hat{T}_{\text{K-G}}(\hat{x},p)$, one has

$$\hat{T}_{\text{K-G}}(\hat{x},p) = -i[\frac{\partial}{\partial E} - \frac{m^2}{2E(E^2 - m^2)}], \quad (19)$$

which implies that the canonical commutation relation $[\hat{T},H] = -i$. Furthermore, one can simplify Eq. (19) by passing from the momentum representation to the energy representation, the relation between the amplitudes for expansions over eigenstates normalized respectively to $\delta(p - p')$ and $\delta(E - E')$ being

$$|p\rangle_{\text{K-G}} \to |E\rangle_{\text{K-G}} = [E^2/(E^2 - m^2)]^{1/4}|p\rangle_{\text{K-G}}, \quad (20)$$

which implies that, in the energy representation, Eq. (19) becomes

$$\hat{T}_{\text{K-G}}(E) = -i\partial/\partial E. \quad (21)$$

Now, let us consider the free Dirac particles in (1+1) dimensional space-time. The time-of-arrival operator $\hat{T}_{\text{Dirac}}$ is given by Eq. (17) or (18), with the Hamiltonian being

$$\hat{H} = \alpha_1 \hat{p} + \beta m, \text{ or } H = \alpha_1 p + \beta m, \tag{22}$$

where $\alpha_1$ and $\beta$ are the Dirac matrices satisfying $\alpha_1^2 = \beta^2 = 1$ and $\alpha_1\beta + \beta\alpha_1 = 0$. One can easily examine the canonical commutation relation $[\hat{T}_{\text{Dirac}}, \hat{H}] = -i$. Substituting Eq. (22) into Eq. (17), one has

$$\hat{T}_{\text{Dirac}}(x, \hat{p}) = -\alpha_1 x - \beta \hat{\tau}, \tag{23}$$

where

$$\hat{\tau} = -\hat{T}_{\text{non}}(x, \hat{p}) = m(\hat{p}^{-1}x + x\hat{p}^{-1})/2 \tag{24}$$

is called the proper time-of-arrival operator. In fact, using the classical expressions for the relativistic time-of-arrival $t = -xE/p$ one has $t^2 - x^2 = (xm/p)^2 = \tau^2$, and then the nonrelativistic time-of-arrival $\pm xm/p$ plays the role of proper time-of-arrival. By quantization, Eq. (24) represents the nonrelativistic time-of-arrival operator when a particle with the momentum $p$, initially at the origin and passes point $x$, and then $\hat{\tau}$ also represents the proper time-of-arrival operator, its classical expression is $\tau = xm/p$. It is interesting to note that, the time operator $\hat{T}_{\text{Dirac}} = -\alpha_1\hat{x} - \beta\hat{\tau}$ is to $t^2 = x^2 + \tau^2$, as the Hamiltonian $\hat{H} = \alpha_1\hat{p} + \beta m$ is to $E^2 = p^2 + m^2$, which shows us a duality between the coordinate space and the momentum space.

## 5. Time operator as the restriction of energy shift generator

Physical observables involve energy differences and not the absolute value of the energy, and do not depend on the choice of zero-energy reference point. To display this kind of physical property, let us consider a system described by the Schrödinger equation

$$i\frac{\partial}{\partial t}|t\rangle = \hat{H}|t\rangle, \tag{25}$$

where $\hat{H}$ stands for the Hamiltonian. Let $\mathcal{R}$ denote the set of real numbers (also called the full real line), a c-number *energy parameter* $e \in \mathcal{R}$ is introduced, which has the dimension of energy and does not depend explicitly on the space-time coordinate $x^\mu = (t, \boldsymbol{x})$: $\partial e / \partial x^\mu = 0$ ($\mu = 0, 1, 2, 3$). In terms of the energy parameter $e$ rewriting

$$\hat{H} = \hat{H}(e) = \hat{H}(0) + e, \ |t\rangle = |t, 0\rangle, \tag{26}$$

Eq. (25) becomes

$$i\frac{\partial}{\partial t}|t, 0\rangle = \hat{H}(e)|t, 0\rangle, \tag{27}$$

Assume that the spectrum of $\hat{H}(0)$ is $\mathcal{R}^+ = \{E \mid 0 \le E < +\infty\}$, as for a system with the Hamiltonian $\hat{H}(e) = \hat{H}(0) + e$, its zero-energy reference point is referred to as $e$, because the spectrum of $\hat{H}(e)$ is $\{E \mid e \le E < +\infty\}$. By rewriting Eq. (25) as Eq. (27), the *e*-dependence, i.e., the dependence of the zero-energy reference point, is carried by the Hamiltonian $\hat{H} = \hat{H}(e)$, and the degree of freedom related to the zero-energy reference point is displayed explicitly. For each given $e$, the spectrum of the Hamiltonian $\hat{H}(e)$ is always bounded from below.

Let $\mathcal{H}_e$ denote the Hilbert space spanned by the eigenstates of $\hat{H}(e)$, $\mathcal{H}$ denote the union of $\mathcal{H}_e$ for all $e \in \mathcal{R}$:

$$\mathcal{H} = \bigcup_{e \in \mathcal{R}} \mathcal{H}_e, \tag{28}$$

which is called the whole Hilbert space. In general, $\mathcal{H}_e$ is the Hilbert space of square integrable functions on $[e, +\infty)$, while $\mathcal{H}$ is the Hilbert space of square integrable functions on $(-\infty, +\infty)$. Moreover, the set of $\mathcal{H}_e$, i.e., { $\mathcal{H}_e$, $e \in \mathcal{R}$ } form a lattice [30].

As mentioned above, by Eq. (27) the Hamiltonian $\hat{H} = \hat{H}(e)$ carries the *e*-dependence. Another alternative is to peel off the *e*-dependence from the operator and associate it with the state by the following equation:

$$\langle t_1, 0 | \hat{H}(e) | t_2, 0 \rangle = \langle t_1, e | \hat{H}(0) | t_2, e \rangle, \tag{29}$$

where

$$\begin{cases} \hat{H}(e) = \hat{V}^+(e,0)\hat{H}(0)\hat{V}(e,0) \\ |t,e\rangle \equiv \hat{V}(e,0)|t,0\rangle \\ \hat{V}(e,0) \equiv \exp(ie\hat{S}) = \exp[i(e-0)\hat{S}] \end{cases}, \tag{30}$$

where $\hat{V}^+(e,0) = \exp(-ie\hat{S})$ stand for the hermitian conjugate of $\hat{V}(e,0)$, $\hat{V}(e,0)$ is called energy shift operator, $\hat{S}$ is the generator of $\hat{V}(e,0)$. To guarantee that $\hat{V}(e,0)$ be a unitary operator and stand for a symmetry transformation in physics, the domain of $\hat{S}$, $\mathcal{D}(\hat{S})$ say, should be chosen as a dense domain of $\mathcal{H}$, which for convenience, is denoted as (and so on)

$$\mathcal{D}(\hat{S}) = \mathcal{H} = \bigcup_{e \in \mathcal{R}} \mathcal{H}_e. \tag{31}$$

For the moment the energy-shift generator $\hat{S}$ is self-adjoint. Using Eqs. (26), (29) and (30), one can obtain the commutation relation that is valid in any closed subset of $\mathcal{D}(\hat{S})$

$$[\hat{H}(0), \hat{S}] = [\hat{H}(e), \hat{S}] = -i. \tag{32}$$

In terms of $|t,0\rangle$ and $\hat{H}(e)$ (and so on) one establishes a *generalized* Heisenberg picture, while in terms of $|t,e\rangle$ and $\hat{H}(0)$ (and so on) one works in a *generalized* Schrödinger picture. Therefore, one has two equivalent approaches to describe the dependence of the zero-energy reference point: choosing a new zero-energy reference-point, in the generalized Heisenberg picture one may keep the state vectors unchanged, while in the generalized Schrödinger picture one may keep the physical operators unchanged. By Eq. (30) the two pictures are related to each other by the unitary operator $\hat{V}(e,0)$.

Using Eq. (30) one can easily show that, in the generalized Schrödinger picture, Eq. (27)

becomes

$$i\frac{\partial}{\partial t}|t,e\rangle = \hat{H}(0)|t,e\rangle. \tag{33}$$

In general, starting from an operator $\hat{F}(e) = \hat{V}^+(e,0)\hat{F}(0)\hat{V}(e,0)$ with $\partial \hat{F}(0)/\partial e = 0$, one can easily obtain, in the generalized Heisenberg picture

$$\frac{d\hat{F}(e)}{de} = i[\hat{F}(e), \hat{S}], \tag{34}$$

Eq. (34) is called the Heisenberg-like equation. Likewise, in the generalized Schrödinger picture, one can easily obtain

$$-i\frac{\partial}{\partial e}|t,e\rangle = \hat{S}|t,e\rangle, \tag{35}$$

Eq. (35) is called the Schrödinger-like equation, which is also referred to as describing the energy shift of states of a system.

Furthermore, if a physical quantity $M$ satisfies $\partial M/\partial e = 0$, $M$ is called a conserved quantity with respect to $e$ (i.e., $M$ has a value constant in $e$), which implies that $M$ is invariant under the energy-shift transformation. For example, let $\{E_n\}$ ($n=0,1,2…$) stand for an energy level structure, then the energy level interval $\Delta E_{mn} = E_m - E_n$ ($m,n = 0,1,2,...$, $m \neq n$) is invariant under the energy-shift transformation. Correspondingly, under the energy-shift transformation, the invariance of a system implies the independence of zero-energy reference point, i.e., the energy-shift symmetry.

Now, let us consider the eigenequation:

$$\hat{H}(0)|E,0\rangle = E|E,0\rangle, (E \in [0,+\infty)). \tag{36}$$

Here for convenience the eigenstate $|E\rangle$ is rewritten as $|E,0\rangle$ (note that it is related to $|t,e\rangle$ rather than to $|t,0\rangle$). In the generalized Schrödinger picture, using the energy-shift operator $\hat{V}(e,0) = \exp(ie\hat{S})$ one makes a unitary transformation

$$|E\rangle = |E,0\rangle \to |E,e\rangle \equiv \exp(ie\hat{S})|E,0\rangle, \, (e \in \mathcal{R}). \tag{37}$$

Owing to the commutation relation $[\hat{H}(0), \exp(ie\hat{S})] = e\exp(ie\hat{S})$, one has

$$\hat{H}(0)|E,e\rangle = (E+e)|E,e\rangle, \, (E \in [0,+\infty)). \tag{38}$$

That is, the new state $|E,e\rangle$ is also the eigenstate of the Hamiltonian $\hat{H}(0)$ with the eigenvalue $(E+e)$. Comparing Eq. (38) with Eq. (36) one has $|E,e\rangle = |E+e,0\rangle$, Eq. (38) can be rewritten as

$$\hat{H}(0)|E+e,0\rangle = (E+e)|E+e,0\rangle, \, ((E+e) \in [e,+\infty)). \tag{39}$$

On the other hand, in the generalized Heisenberg picture, an equivalent description for the transformation (37) can be obtained, where the state vector $|E\rangle = |E,0\rangle$ keeps unchanged, while the Hamiltonian operator transforms as follows

$$\hat{H}(0) \to \hat{H}(e) = \exp(-ie\hat{S})\hat{H}(0)\exp(ie\hat{S}) = \hat{H}(0) + e, \, (e \in \mathcal{R}). \tag{40}$$

Because of $[\hat{S},\hat{S}] = 0$, the generator $\hat{S}$ keeps invariant: $\hat{S}(0) \to \hat{S}(e) = \hat{S}(0)$. Therefore, in the generalized Heisenberg picture Eq. (38) or (39) becomes

$$\hat{H}(e)|E,0\rangle = [\hat{H}(0) + e]|E,0\rangle = (E+e)|E,0\rangle, \, (E \in [0,+\infty)). \tag{41}$$

Comparing Eq. (41) with Eq. (36), one can show that, the transformation (37) or (40) is equivalent to putting the system with the Hamiltonian $\hat{H}(0)$ into a constant and uniform potential field with the potential of $e$, such that the new Hamiltonian is $\hat{H}(e) = \hat{H}(0) + e$. In other words, the transformation from Eq.(36) to (41) is identical with changing the zero-energy reference point of the system from 0 to $e$. The equivalence between Eqs. (41) and (36) shows us the energy-shift symmetry, i.e., the independence of the zero-energy reference points.

It is very important to note that, as shown in Eq. (38) or (39), on the one hand, for all $e \in \mathcal{R}$, all possible eigenstates of $\hat{H}(0)$ span the whole Hilbert space $\mathcal{H} = \bigcup_{e \in \mathcal{R}} \mathcal{H}_e$; on the other hand, once a zero-energy reference point $e$ is given, the Hilbert space of the system is given by $\mathcal{H}_e = \{|E, e\rangle\}$, for the moment the spectrum of $\hat{H}(e) = \hat{H}(0) + e$ is bounded from below (i.e., $(E + e) \in [e, +\infty)$). Therefore, only a part of eigenstates of $\hat{H}(0)$ represent the physical states of the system, they span the Hilbert space $\mathcal{H}_e = \{|E, e\rangle\}$ for the given zero-energy reference point $e$, and in $\mathcal{H}_e = \{|E, e\rangle\}$ the Hamiltonian spectrum is bounded from below; while the other eigenstates of $\hat{H}(0)$ stand for the unphysical states, they belong to the Hilbert space $(\mathcal{H} - \mathcal{H}_e)$ and are valid only in the sense of mathematics (mathematically all coordinates and momentums included in the Hamiltonian can be any complex number, such that the Hamiltonian spectrum can be unbounded in mathematics). The existence of the unphysical states presents no problem: the transition probability between the physical and unphysical states vanishes, which we will discuss later.

As mentioned before, choosing a given zero-energy reference point $e \in \mathcal{R}$, the physical system described by Eq. (38) or (39) is fixed on, and its Hilbert space is $\mathcal{H}_e = \{|E, e\rangle\}$. On the other hand, the domain of the energy-shift generator $\hat{S}$ is a dense domain of $\mathcal{H} = \bigcup_{e \in \mathcal{R}} \mathcal{H}_e$. Due to Eq. (32), the restriction of $\hat{S}$ to the Hilbert space $\mathcal{H}_e$, denoted as $\hat{T}$, satisfies

$$[\hat{H}(0), \hat{T}] = [\hat{H}(e), \hat{T}] = -\mathrm{i}. \tag{42}$$

For the moment, $\mathcal{D}(\hat{H}(0)) = \mathcal{D}(\hat{T}) = \mathcal{H}_e$. Eq. (42) implies that $\hat{T}$ represents the time operator of the system. Therefore, the time operator is the restriction of the energy-shift generator $\hat{S}$ to the Hilbert space of the system. Obviously $\hat{T}(e) = \hat{T}(0)$ because of $\hat{S}(e) = \hat{S}(0)$, i.e., they are independent of zero-energy reference point.

It is easy to prove that, the time operator $\hat{T}$ is not a self-adjoint operator, which due to the fact that, in the Hilbert space $\mathcal{H}_e$, the spectrum of the Hamiltonian is bounded from below (a related discussion can see Ref. [11], for example). As a result, the restriction of the energy-shift operator $\hat{V}(e,0) = \exp(ie\hat{S})$ to $\mathcal{H}_e$, i.e., $\exp(ie\hat{T})$, is no longer a unitary operator. For the moment, Pauli's objection is not valid. On the other hand, likewise, using the transformation $|E,0\rangle \to |E,e\rangle = \exp(ie\hat{T})|E,0\rangle$, ($e \in \mathcal{R}$), one can obtian all possible eigenstates of $\hat{H}(0)$, where only a part of them represent the physical states, while the others are valid only in the sense of mathematics.

Now, let us prove that the transition probability between the physical states and unphysical states vanishes. For convenience, let $\mathcal{H}_0 = \{|E,0\rangle\}$ be the Hilbert space of a system, and the orthonormality and completeness relations are given by

$$\begin{cases} \int_0^{+\infty} dE |E,0\rangle\langle E,0| = 1 \\ \langle E,0|E',0\rangle = \delta(E - E') \end{cases}, \quad E, E' \in [0, +\infty). \tag{43}$$

Let

$$C(E', E, e) \equiv \langle E', 0|E, e\rangle, \tag{44}$$

where $|E,e\rangle$ is obtained by $|E,0\rangle \to |E,e\rangle = \exp(ie\hat{T})|E,0\rangle$, $e \in \mathcal{R}$. Using Eqs. (43)-(44) one has

$$|E,e\rangle = \int_0^{+\infty} dE_\lambda C(E_\lambda, E, e)|E_\lambda, 0\rangle. \tag{45}$$

Consider that

$$\begin{aligned}\langle E', 0|\hat{H}(0)|E, e\rangle &= \langle E', 0|(E + e)|E, e\rangle \\ &= (E + e)C(E', E, e)\end{aligned}, \tag{46}$$

$$\begin{aligned}\langle E', 0|\hat{H}(0)|E, e\rangle &= \langle E', 0|\hat{H}(0)\int_0^{+\infty} dE_\lambda C(E_\lambda, E, e)|E_\lambda, 0\rangle \\ &= E' C(E', E, e)\end{aligned}, \tag{47}$$

one has

$$(E+e)C(E',E,e) = E'C(E',E,e), (e \in \mathcal{R}).  \quad (48)$$

Applying Eqs. (36)-(39) one can discuss Eq. (48) as follows:

1). If $(E+e) \in [0,+\infty)$, one has $|E,e\rangle = |E+e,0\rangle \in \mathcal{H}_0 = \{|E,0\rangle\}$, i.e., $|E,e\rangle$ is also a physical state of the system. Using Eqs. (43)-(44) and (48) one has

$$C(E',E,e) = \langle E',0|E,e\rangle = \delta(E'-E-e). \quad (49)$$

2). If $(E+e) \notin [0,+\infty)$, because $E' \in [0,+\infty)$ one has $E+e \neq E'$, then Eq. (48) implies that

$$C(E',E,e) = \langle E',0|E,e\rangle \equiv 0. \quad (50)$$

On the other hand, the condition $(E+e) \notin [0,+\infty)$ implies that $|E,e\rangle$ is no longer a physical state of the system: $|E,e\rangle = |E+e,0\rangle \notin \mathcal{H}_0 = \{|E,0\rangle\}$. Therefore, Eq. (50) implies that the transition probability ($\propto |C(E',E,e)|^2$) between the physical state $|E,0\rangle$ and the unphysical state $|E,e\rangle$ ($e < -E$) vanishes. According to the time-dependent perturbation theory of quantum mechanics, all the perturbed states of a system can be expressed as a linear combination of the physical states of the system, then the transition probabilities between the perturbed states and the unphysical states also vanish. Therefore, the time operator $\hat{T}$, as the restriction of the energy-shift generator $\hat{S}$ to the Hilbert space of a physical system, can be introduced without presenting any problem.

## 6. Discussions and conclusions

Up to now, we have shown that time operator can be introduced by three different approaches.

By pertaining time operator to dynamical variables, one either preserves the requirement that time operator be conjugate to the Hamiltonian but abandons the self-adjointness of time operator, or maintains the self-adjointness of time operator, but the canonical commutation relation between time operator and the Hamiltonian is valid only in the sense of quantum-mechanical average. This method is similar to the Mandelstam–Tamm version of the time-energy uncertainty.

By quantizing the classical expression of time, one can obtain a time operator. That is, the transition from the classical expression to a quantum-mechanical description requires us to symmetrize the classical expression and replace all dynamical variables with the corresponding linear operators. However, this method is not valid for tunneling time, because quantum tunneling is a purely quantum-mechanical effect, and there does not exist any classical expression for tunneling time.

The method that taking time operator as the restriction of energy shift generator is based on the fact that: physical observables involve energy differences and not the absolute value of the energy, which implies that physical laws do not depend on the choice of zero-energy reference point. The energy-shift operator is defined on a dense domain of the Hilbert space of square integrable functions on the full real-line. On the other hand, as the Hamiltonian spectrum of a physical system is bounded from below, the time operator of the system must be defined on a dense domain of the Hilbert space of square integrable functions on the half real-line. As a result, the energy-shift generator is a Naimark's dilation of the time operator, while the time operator corresponds to the restriction of the energy-shift generator to the Hilbert space of the system, and it is not a self-adjoint operator. However, according to the formalism of POVMs, the time operator can still represent an observable. Therefore, the traditional concept of observables can be extended in such a way: if an operator is self-adjoint in a Hilbert space, then its restriction to a subspace of the Hilbert space can represent an observable (even if it is no longer a self-adjoint operator).

**Acknowledgements**

The first author (Z. Y. Wang) would like to greatly thank Dr. Changhai Lu and Prof. J. G. Muga for their many helpful discussions and valuable suggestions. Project supported by the Specialized Research Fund for the Doctoral Program of Higher Education of China (Grant No. 20050614022) and by the National Natural Science Foundation of China (Grant No. 60671030).

**References**

[1] M. Büttiker and R. Landauer, Phys. Rev. Lett. 49 (1982) 1739-1742.

[2] H. A. Fertig, Phys. Rev. Lett. 65 (1990) 2321-2324.

[3] R. Landauer and Th. Martin, Rev. Mod. Phys. 66 (1994) 217-228.

[4] A. M. Steinberg, Phys. Rev. Lett. 74 (1995) 2405-2409.

[5] Ph. Balcou and L. Dutriaux, Phys. Rev. Lett. 78 (1997) 851-854.

[6] A. L. Xavier, Jr. and M. A. M. de Aguiar, Phys. Rev. Lett. 79 (1997) 3323-3326.

[7] Q. Niu and M. G. Raizen, Phys. Rev. Lett. 80 (1998) 3491-3494.

[8] N. Yamada, Phys. Rev. Lett. 83 (1999) 3350-3353.

[9] P. Pereyra, Phys. Rev. Lett. 84 (2000) 1772-1775.

[10] F. Grossmann, Phys. Rev. Lett. 85 (2000) 903-907.

[11] J. G. Muga and C. R. Leavens, Phys. Rep. 338 (2000) 353-438.

[12] H. G. Winful, Phys. Rev. Lett. 91 (2003) 260401.

[13] Z. Y. Wang, B. Chen and C. D. Xiong, J. Phys. A: Math. Gen. 36 (2003) 5135-5147.

[14] N. Yamada, Phys. Rev. Lett. 93 (2004) 170401.

[15] V. S. Olkhovsky, E. Recami and J. Jakiel, Phys. Reports 398 (2004) 133-178.

[16] H. G. Winful, M. Ngom and N. M. Litchinitser, Phys. Rev. A 70 (2004) 052112.

[17] A. Ranfagni, I. Cacciari and P. Sandri *et al*, Phys. Lett. A 343 (2005) 469-473.

[18] A. K. Pan, Md. M. Ali and D. Home, Phys. Lett. A 352 (2006) 296-303.

[19] J. G. Muga, R. Sala, and I. L. Egusguiza, in Time in Quantum Mechanics, Springer, Verlag Berlin Heidelberg, 2002.

[20] W. Pauli, in Encylopaedia of physics, edited by S. Flugge (Springer, Berlin, 1958), Vol. 5, p. 60.

[21] J. G. Muga, C. R. Leavens, and J. P. Palao, Phys. Rev. A 58 (1998) 4336-4344.

[22] I. L. Egusquiza and J. G. Muga, Phys. Rev. A 61 (1999) 012104.

[23] E.A. Galapon, R. Caballar, and R. Bahague, Phys. Rev. Lett. 93 (2004) 180406.

[24] M. Skulimowski, Phys. Lett. A 297 (2002) 129-136.

[25] M. Skulimowski, Phys. Lett. A 301 (2002) 361-368.

[26] R. Giannitrapani, Int. J. Theor. Phys. 36 (1997) 1575-1587.

[27] L. Mandelstam and I. Tamm, J. Phys. (SSSR) 9 (1945) 249-254.

[28] L. H. Ryder, Quantum Field Theory (2nd ed), Cambridge University Press, England, 2003, pp 146-150.

[29] P. Busch, Operational Quantum Physics, Springer-Verlag, Berlin, 1995, p. 81.

[30] J. M. Jauch, Foundations of Quantum Mechanics, Addison-Wesley Publishing Company, Inc. Philippines, 1968, p.26.